\documentclass[11pt,twoside]{article}
\usepackage{asp2006_hotcool}
\usepackage{epsf}
\usepackage{graphicx}
\begin{document}
\title{The effects of red supergiant mass loss on\\ supernova ejecta and the
circumburst medium}
\author{Jacco Th. van Loon}
\affil{Lennard-Jones Laboratories, Keele University, Staffordshire ST5 5BG,
United Kingdom}
\begin{abstract}
Massive stars becoming red supergiants lose a significant amount of their mass
during that brief evolutionary phase. They then either explode as a
hydrogen-rich supernova (SN Type II), or continue to evolve as a hotter
supergiant (before exploding). The slow, dusty ejecta of the red supergiant
will be over-run by the hot star wind and/or SN ejecta. I will present
estimates of the conditions for this interaction and discuss some of the
implications.
\end{abstract}

\section{Red supergiants}

At first glance the definition of a red supergiant seems clear-cut: it is a
large star with a red colour. The red colour is a result of the photosphere
being cool, which in turn is the result of the adjustment of the structure of
the stellar mantle to facilitate energy transport (which through a large
portion of the mantle happens {\it via} convection) from the stellar interior
where it is produced out to the photosphere. For a cool star to radiate away
the energy at the same high rate at which it is produced by nuclear burning,
the radiating surface must be enormous, as $L_\star=4\pi R_\star^2\sigma
T_{\rm eff}^4$. The {\it super} in supergiant refers to the effective gravity
being particularly low; the effect of the effective gravity on the shapes of
spectral lines forms the basis for the Morgan-Keenan system of luminosity
classes. In practice, this often (but not always) translates into a higher
luminosity for the red supergiants, which have luminosity class I, than for
the {\it not-so-super} red giants, which have luminosity classes II (high up
the Asymptotic Giant Branch) or III (lower on the AGB, or on the first-ascent
Red Giant Branch). The specific nuclear burning rate increases with increasing
pressure and temperature, and the luminosity further increases with increasing
mass (fuel) of the nuclear engine; these all increase in more massive stars,
and thus red supergiants are features of the evolution of massive stars. The
underlying reasons for the change from a relatively compact main-sequence star
to a red supergiant remain elusive --- it is not merely due to their huge
luminosity, as the evolution of massive stars towards (and away from) the red
supergiant phase is accompanied by relatively little change in luminosity.

At second glance, it is not obvious at what point a star can be called a red
supergiant. How cool must it be? Stars with spectral types similar to that of
the Sun, but luminosity class I are called yellow supergiants, or yellow
hypergiants to distinguish them from the luminosity class II post-AGB stars.
Very luminous K-type giants could be called orange supergiants. It seems
logical to define red supergiants as those stars of luminosity class I and an
M-type spectrum. This turns out to be a very useful criterion, as the M class
is characterised by the appearance of additional molecular absorption bands in
the optical spectrum, and this is accompanied (if not followed) by other
important phenomena such as radial pulsation and circumstellar dust. The other
criterion, the luminosity class, is complicated by the fact that at the lower
end of the mass spectrum of red supergiant progenitors, red supergiants are
not necessarily more luminous than stars at the tip of their AGB, descendant
from relatively massive intermediate-mass stars. If this is not enough, these
red supergiants are also warmer and thus smaller than their AGB
luminosity-equivalents; and the most extreme AGB stars will have reduced their
mass to close of that of a white dwarf (which will have further contributed to
their increase in size and reduction in photospheric temperature), while red
supergiants may still carry a significant fraction of their birth mass.
Consequently, the effective gravity of a red supergiant can be much higher
than that of a red giant.

For the purpose of this discourse, and as a suggestion for a meaningful use of
the term, we introduce here a hybrid definition of red supergiant, marrying an
observational with a theoretical criterion: {\it a red supergiant is a massive
star with spectral type M}. Note that we have not yet specified what we mean
by massive.

\subsection{The nature of red supergiant progenitors and descendants}

So which stars make it to the red supergiant phase? Often, a lower birth mass
of $M_{\rm birth}>8$ M$_\odot$ is adopted for stars that upon exhaustion of
helium-burning, and contrary to AGB stars, proceed to ignite carbon in their
cores. But this relies on the treatment of convection and other types of
mixing in the core and circum-core regions, and can be as low as 5--6
M$_\odot$ in models with efficient convective overshoot (the ability of
convection cells to penetrate regions that are formally stable against
convection, due to inertia). The stars that straddle the distinction between
entrance into the red supergiant phase or termination on the AGB have ages of
around 40 Myr. This is a few times longer than typical lifetimes of molecular
clouds, and at a modest peculiar velocity of 1 km s$^{-1}$ the star would have
wandered off by 40 pc anyhow; red supergiants (and their immediate
progenitors) can thus be encountered well outside H\,{\sc ii} regions.

Stars too massive do not become red supergiants. They terminate their
evolution prematurely, as Wolf-Rayet stars with helium-rich photospheres, or
as Luminous Blue Variables hitting the S\,Dor instability strip (or the
Humphreys-Davidson Eddington-luminosity limit in the case of the most massive
stars). The transition regime between stars that do become red supergiants and
those that do not, occupies approximately the $M_{\rm birth}=25$--30 M$_\odot$
range. However, it appears to depend on details that may differ between stars
of the same birth mass: the overall metallicity of a star affects the opacity
profile throughout the stellar mantle, and thus its structure; rotation could
lead to extra mixing of nucleosynthesis products throughout the stellar
interior, thereby facilitating transformation into a red supergiant; mass loss
in the preceding stages also affects the prospects for a star to become a red
supergiant, and this likely depends on metallicity and angular momentum. The
upper mass limit to red supergiant progenitors is difficult to determine
observationally, as massive stars are under-represented in co-eval stellar
aggregates and the red supergiant evolutionary phase is particularly brief.
But the absence of evidence for red supergiants younger than $\sim10$ Myr
supports the afore-mentioned mass cut-off.

Red supergiants have reduced mass, compared to their birth mass. During their
lives on the main sequence, but more importantly --- especially for the
lower-mass stars --- during the post-main sequence phase as blue, B-type
supergiants, the progenitors of red supergiants lose mass through a wind
driven by radiation pressure mainly in the UV transitions of metallic ions.
The mass-loss rates typically reach values of $\dot{M}_{\rm blue}\sim{\rm
few}\times10^{-6}$ M$_\odot$ yr$^{-1}$. In this manner, a few M$_\odot$ may be
lost prior to the red supergiant phase. The mass-loss rate can not be computed
{\it ab initio}, and direct measurements may have been over-estimated by up to
an order of magnitude because of the unknown degree of clumping in the wind.
The subsequent migration across the A--K spectral types is rapid, and mass
loss then thus seems of limited importance. However, LBVs are seen also at
relatively-low luminosity; in fact the S\,Dor instability strip appears to
span across the Hertzsprung-Russell diagram, from the most massive blue
hypergiants down towards, and intersecting if extra-polated, the AGB. Stars
evolving from the main-sequence towards the red supergiant phase inevitably
cross this strip. What determines the instability --- and thus the symptoms
these stars develop --- is unclear. It is also unclear how much additional
mass is shed during the LBV phase (or cycles). There thus remains a
possibility that red supergiants can have mantles of very little mass. This is
important, as this would facilitate the transformation of red supergiants into
Wolf-Rayet stars due to the drastic reduction in mantle mass (and thus the
inwards migration of the photosphere) resulting from mass loss during the red
supergiant phase.

For reasons yet unclear, less massive red supergiants may undertake an
excursion towards hotter photospheric temperatures, possibly to return to
enter the red supergiant phase for a second time. These blue loops resemble in
some respects the horizontal-branch phase that metal-poor low-mass stars
experience in between the RGB and AGB phases. In conclusion, red supergiants
may have had a very rich and complicated live indeed, and likewise hot
supergiants may have gone through a red supergiant phase some time in their
recent past.

\subsection{Supernovae with red supergiant progenitors}

%
\begin{figure}[!ht]
\includegraphics[width=13.3cm]{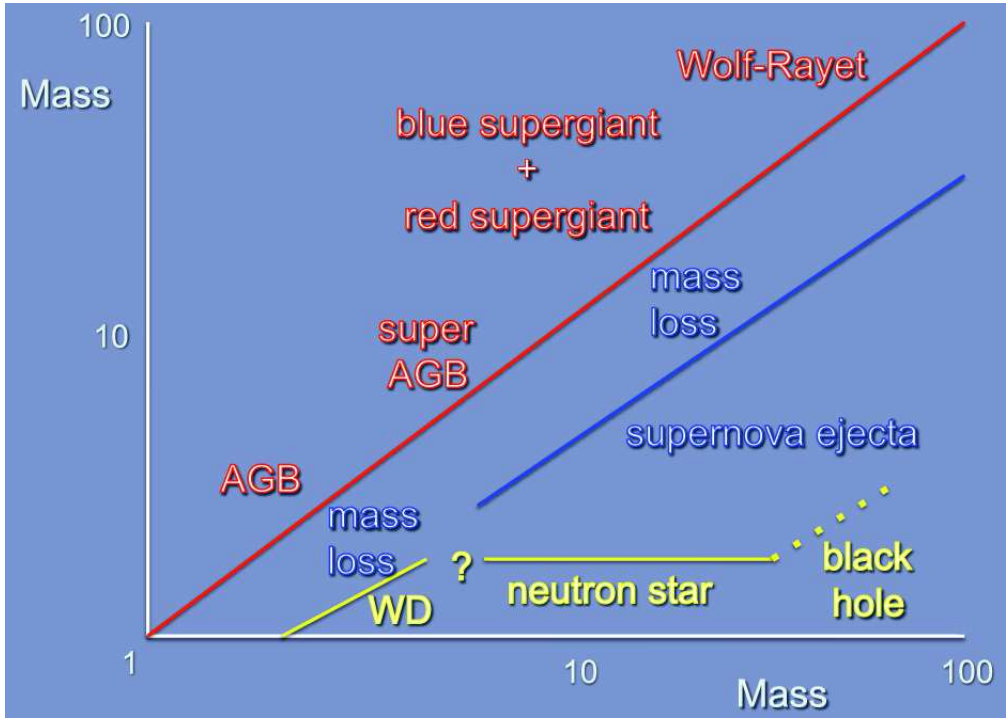}
\caption{Schematic context for red supergiant evolution and their mass loss.}
\end{figure}

Red supergiants are the classical, predicted progenitors of core-collapse
supernovae. Upon helium exhaustion, the star will proceed rapidly through
subsequent successive stages of nuclear fusion until it no longer is an
exothermic process, the core will collapse under its own gravity until halted
by neutron pressure, the inner mantle will bounce off the newly formed neutron
star, and the resulting shock wave will traverse the outer mantle which had
hitherto been unaware of the dramatic events unfolding inside. Thus a
hydrogen-rich (Type II) supernova event occurs. If the red supergiant
transformed into a Wolf-Rayet star before exploding, the supernova would be of
a hydrogen-poor type, Type Ib or c (Type Ia is usually linked with white
dwarves). If a black hole formed instead of a neutron star, which might be the
case in massive red supergiants, then the shock wave is likely to be much
weaker (in the absence of a solid surface to bounce off) and thus the
supernova fainter, and possibly the ejecta would be smaller in mass. On the
other hand, the relatively light mantles of the least massive red supergiants
might be entirely shed during their relatively long-lasting red supergiant
phase. Like other red supergiants, these stars go on to burn carbon in their
cores, but like AGB stars they experience thermal pulses as different shells
burn alternately. These so-called super-AGB stars might terminate their
evolution prematurely, leaving behind an oxygen-neon white dwarf; however, if
the mass loss is insufficient they will proceed to the next stage, and in
their case electron capture will cause a fall in pressure which induces a
supernova event, probably of Type II. Figure 1 presents a schematic overview
--- the boundaries in mass between the various scenarios, and between the mass
lost through pre-supernova mass loss and supernova ejecta, are all rather
insecure.

Red supergiants have recently been confirmed directly to be associated with
supernovae of Type II-P (Smartt et al.\ 2009). These supernovae have
lightcurves characterised by a plateau phase (hence the P sub-type); which is
when the ejecta have large opacity due to their ionisation by the shock wave,
causing the storage of photons within the expanding ejecta until the ejecta
recombine, the opacity drops, and the photons escape freely. Clearly, these
red supergiants were some way from shedding their mantle through (more gentle)
mass loss. The masses of these red supergiant supernova progenitors span a
range in mass of $M_{\rm birth}=8.5$--16.5 M$_\odot$ (Smartt et al.\ 2009).
This leaves little room for super-AGB stars to yield a markedly different type
of supernova from their more massive siblings. It has been suggested that the
core collapse in the most massive red supergiants, in the mass range of 17--25
(or higher) M$_\odot$, might result in black holes and possibly lead to faint
events of Type II-L (with optically thin ejecta causing a linear decay in the
lightcurve, lacking the plateau).

\section{Red supergiant mass loss}

\subsection{Characteristics of red supergiant winds}

In his seminal paper on the discovery of a wind emanating from the M5-type red
supergiant $\alpha$\,Herculis, the massive component of a visual binary, Armin
Deutsch (1956) laid the foundations for what we know about red supergiant mass
loss. The optical, violet-displaced absorption lines were used to infer a wind
speed of $v_{\rm wind}\sim10$ km s$^{-1}$, and a mass-loss rate of at least
$\dot{M}>3\times10^{-8}$ M$_\odot$ yr$^{-1}$. He also pointed out that the
line broadening in the photospheres of red supergiants suggests turbulent
motions at the level of a few km s$^{-1}$. One might expect such turbulence to
persist through at least the inner regions of the outflow. He also noted the
irregular variability of the light output from the star; the associated radial
motions at the base of the atmosphere also reach a few km s$^{-1}$. Thus, the
slow winds from cool supergiants are not expected to be homogeneous. Indeed,
Deutsch inferred a clumpy medium, with electron densities $n_{\rm e}\sim10^6$
cm$^{-3}$ inside clumps and a volume filling factor of only 1:$10^7$. Deutsch
realised that this wind is unlikely to be a peculiarity of $\alpha$\,Herculis,
and that in fact mass loss must be commonplace among all M-type giants and
supergiants.

In 1971, Gehrz \& Woolf published another seminal paper, on a sample of M-type
giants and supergiants including $\alpha$\,Herculis: they found that the winds
of these stars are dusty, causing extra emission at mid-IR wavelengths at the
expense of optical light. Adopting a density profile and a gas:dust mass ratio
assuming all silicon were in solid form, they derived a mass-loss rate for
$\alpha$\,Herculis of $\dot{M}\sim9\times10^{-8}$ M$_\odot$ yr$^{-1}$, in
fantastic agreement with the estimate by Deutsch. They realised that these
estimates should be regarded as lower limits, as the dust condensation process
may not be 100\% complete.

How typical is $\alpha$\,Herculis? Gehrz \& Woolf (1971) estimated much higher
rates of mass loss for most other red supergiants, up to $\dot{M}\sim10^{-5}$
M$_\odot$ yr$^{-1}$; in fact, the rates from the dustiest-known red
supergiants --- {\it e.g.}, VY\,Canis\,Majoris and NML\,Cygni --- reach values
of $\dot{M}\sim10^{-4}$ M$_\odot$ yr$^{-1}$ (Jura \& Kleinmann 1990). Wind
speeds of red supergiants are also higher in general than that in
$\alpha$\,Herculis, typically $v_{\rm wind}\sim20$--30 km s$^{-1}$ in the
Milky Way ({\it c.f.}\ Jura \& Kleinmann 1990). The mass-loss rates do not
appear to depend on the metallicity at birth, but the dust:gas mass ratio,
$\psi$ scales in direct proportion to metallicity and the wind speed scales as
$v_{\rm wind}\propto\psi^{0.5}$ (Marshall et al.\ 2004; van Loon et al.\
2005, 2008). In the Milky Way, $\psi\sim1$:200 in relatively dusty stars.

The density in the wind, $\rho=\dot{M}/(4\pi r^2v_{\rm wind})$ spans a vast
range: near the base of the wind, at $r_{\rm in}\sim10^9$ km (several AU), the
density can exceed $\rho\sim10^{-15}$ g cm$^{-3}$ (assuming
$\dot{M}\sim10^{-5}$ M$_\odot$ yr$^{-1}$), or $n\sim10^{12}$ particles
cm$^{-3}$. Assuming the wind was maintained for a period of $10^5$ yr, the
earliest ejecta could possibly have reached a distance of $r_{\rm
out}\sim10^{14}$ km (a few pc); by then, the density would have dropped to
$n\sim100$ particles cm$^{-3}$, or $n\sim1$ particles cm$^{-3}$ for a more
moderate mass-loss rate of $\dot{M}\sim10^{-7}$ M$_\odot$ yr$^{-1}$. This
approaches the density of the diffuse interstellar medium in the Galactic
Disc. The wind cools, {\it via} emission by grains, molecules and atomic
fine-structure lines, and the temperature in the outer regions of the wind
will reach a balance set by the local interstellar radiation field, typically
a few dozen K. With $P_{\rm out}=nkT\sim10^{-20}$ to $10^{-18}$ bar, the
internal pressure of the outer wind may thus not be too different from that of
the interstellar medium. The ram pressure exerted by the bulk flow, $P_{\rm
ram}=\rho v^2\sim10^{-17}$ to $10^{-15}$ bar, {\it i.e.}\ a thousand times
higher than the internal pressure. Close to the star, the wind temperature can
be $T_{\rm in}\sim1000$ K and the internal pressure can be more comparable to
the ram pressure, especially in the relatively-slow winds of metal-poor red
supergiants.

\subsection{The red supergiant mass-loss mechanism}

Deutsch (in his 1956 paper) realised that the wind speed in $\alpha$\,Herculis
is much lower than the escape velocity from the stellar photosphere, $v_{\rm
esc}\sim100$ km~s$^{-1}$. He assessed the possibility that radiation pressure
on the circumstellar ions (and atoms) could drive the wind until the local
escape velocity would have dropped below the wind speed, but the values for
the achieved acceleration he computed fell short by orders of magnitude and
thus he invoked an as yet unidentified mechanism. Gehrz \& Woolf (in their
1971 paper) demonstrated that radiation pressure on dust grains, which have
large continuum opacity (thus easily building a large optical depth, $\tau$)
at optical wavelengths, may be sufficient to drive the wind. Even though the
grains form a minority compound, collisions between grains and gas particles
in the dense inner parts of the winds are frequently enough for the gas to be
dragged along with the grains, and the luminosity of red supergiants is
sufficiently high to provide the required radiation pressure, $P_{\rm
rad}=\tau (L/c)/(4\pi r^2)$, at least until both fluids are gravitationally
unbound from the star (see, {\it e.g.}, Ferrarotti \& Gail 2006). The
dust-driven wind paradigm predicts a dependence of the wind speed on the
luminosity, $v_{\rm wind}\propto L^{0.25}$ (approximately), which has been
confirmed observationally by measurements of OH 1612 MHz masers in the Large
Magellanic Cloud and the Galactic Centre (Marshall et al.\ 2004). The maximum
mass-loss rate attained by red giants and red supergiants appears also to
increase with increasing luminosity (van Loon et al.\ 1999), as well as with
decreasing temperature of the stellar photosphere (van Loon et al.\ 2005).
However, to take this, too, as evidence for a radiation-driven outflow would
be misleading as the paradigm does not, in fact, explain how the wind is
initiated, in the dust-free atmosphere closest to the star. The density at the
critical point, at which acceleration by radiation pressure starts operating,
is set by an alternative mechanism.

The presence of a dusty wind goes hand-in-hand with radial pulsation of the
stellar photosphere. Even though the bolometric amplitude of red supergiants
rarely exceeds a few tenths of a magnitude, compared to a magnitude for dusty
AGB red giants, in terms of absolute units of energy the pulsation of red
supergiants is at least as powerful as that in AGB stars, and a good
correlation is observed between the mass-loss rate and the energy involved in
the pulsation cycle (van Loon et al.\ 2008). However, the energy involved in
the outflow is several orders of magnitude less than that in the pulsation,
and one can not therefore assume that the pulsation directly (mechanically)
drives the mass loss. That said, radial pulsation has two important effects:
it leads to a sustained increase in the scale-height of the atmosphere, and
the associated shocks accommodate the condensation of gas onto nucleation
seeds and further grain growth (Bowen 1988). Relatively-warm red supergiants
(of early-M type) do not always pulsate very strongly, nor possess as much
dust as expected, {\it e.g.}, Betelgeuse (van Loon et al.\ 2005; see also the
contribution by Graham Harper). These stars bear the signature of a
chromospheric temperature-inversion layer; although gas pressure alone is not
thought to be sufficient to drive a Parker-type wind, the magneto-acoustic
waves that are at the origin of heating the chromosphere might dissipate in
the right manner to drive mass loss. The transitions between chromospheric and
pulsation-driven regimes of mass loss, and the associated phase change in the
thermodynamic and compositional state of the wind, are mapped beautifully for
lower-mass red giants by Judge \& Stencel (1991), McDonald \& van Loon (2007),
and McDonald et al.\ (2009).

\subsection{Complications with empirical mass-loss determinations}

One must realise the current limitations on theoretical and empirical values
and formulae for the mass-loss rates of red supergiants. As for theory: no
{\it ab initio} models exist that reliably predict mass-loss rates.
Measurements of mass-loss rates derived from the effects of dust on the
spectral energy distribution (which yields a value for $\tau$) rely on
knowledge of the dust:gas mass ratio, wind speed, and luminosity (hence
distance), parameters which are rarely known accurately. Measurements of
mass-loss rates derived from molecular line emission at micro-wavelengths
suffer from uncertainties in envelope chemistry and again distance, apart from
detecting these lines being more challenging than detecting dust emission. As
already noted by Deutsch (1956), measurements of mass-loss rates from optical
absorption line profiles suffer from uncertainties in the ionization profile
throughout the envelope, as well as the degree of clumping, and from the fact
that they may not trace (only) the mass which eventually escapes. Furthermore,
winds of red supergiants, in particular (or perhaps more easily noticed) the
dustiest ones, are rarely spherically symmetric --- see, {\it e.g.}, the
spectacular resolution of the circumstellar dust envelope surrounding an
extra-galactic red supergiant by Ohnaka et al.\ (2008). Also, the mass-loss
rate of any given star will have varied in the past, and depending on the
method employed a rate will be obtained which refers to a certain moment in
the past (or an average rate, possibly weighted over time in a complicated
manner) --- even the freshest dust was formed from material which left the
stellar photosphere a few years earlier, and cold dust and molecular line
emission may trace mass loss $>10^4$~yr in the past. Correlating these rates
with the current stellar properties may thus be physically inappropriate. To
summarise, individual mass-loss rates of red supergiants are at least a factor
two or so uncertain, but they can be in error by more than an order of
magnitude. Some of the uncertainty may be systematic within a given sample of
stars, which affects the derivation of empirical formulae.

\section{Red supergiant mass loss in evolutionary context}

\subsection{Dependence on the preceding evolutionary phase}

There are two important dependences of red supergiant mass loss on what has
happened before, the first being the structure of the red supergiant mantle,
or: how cool does the photosphere become? This is of crucial importance, as
the mass-loss rate is observed to be higher at a lower temperature (\S 2.2);
this is probably related to the fact that cooler red supergiants have more
tenuous mantles (a million times less dense than their main-sequence mantles!)
which are inherently unstable against fundamental-mode radial pulsation.
Indeed, the closer a star approaches the Hayashi boundary, the closer it is to
hydrostatic imbalance. The intricacies of pre-red supergiant evolution are
more adequately described elsewhere, but we note that, for example, rotation
enhances mixing, causing enhancement of carbon, oxygen, and in particular
nitrogen, throughout the mantle (Meynet, Ekstr\"om \& Maeder 2006). This, in
turn, lowers the photospheric temperature of the red supergiant. It was
suggested that this would also lead to enhanced mass loss, but if it does then
it is more likely to be related to the more unstable mantle, rather than to
enhanced dust formation as a result of a larger amount of refractory elements:
dust {\it condensation} --- in particular in the oxygen-rich environments of
red supergiants --- is probably limited by the abundance of nucleation seeds,
which rely on elements such as titanium, and, with the dominant oxygen-rich
grain material being silicates, grain {\it growth} will be limited by the
availability of silicon (van Loon et al.\ 2008). Neither titanium nor silicon
are enhanced.

Mass loss itself also affects the structure of the mantle. Approximating the
star as a polytrope, the radius of the star will grow as the mass is reduced,
$R\propto M^{-1/3}$. This, in turn, will cool the photosphere (as the
luminosity remains largely unaffected) and possibly enhance the mass loss.
Besides the cooling effect, the growing radius and shrinking mass both lead to
a reduction of the surface gravity of the star; mass-loss rates of red giants
are known to show a clear anti-correlation with surface gravity (Judge \&
Stencel 1991; McDonald et al.\ 2009). It is not understood what would prevent
this from becoming a runaway process --- perhaps this causes the final
ejection of a proto-planetary nebula by an AGB star, but a similar sudden
shell-ejection is not known to occur in red supergiants. The effect of a
larger radius would also be noticed in the length of the pulsation period,
which is relatively straightforward to measure (and which does not depend on
distance), as for a harmonic oscillator, $P\propto R^3$. Hence, $P\propto
M^{-1}$. As the luminosity is a reflection of the birth mass, the difference
between the luminous mass and the pulsational mass, $\Delta M\equiv M_{\rm
lum}-M_{\rm puls}$ measures the time-integrated mass loss. This would be
worthwhile to attempt in the recently-discovered massive, young Galactic
clusters; these harbour more than a dozen red supergiants each (see Ben
Davies' chapter in this book), thus each presenting a series of snapshots out
of the movie telling the life-story of a star. This can then be tied in with
direct measurements of the mass-loss rates along these cluster red supergiant
sequences.

The second dependence is that of the circumstellar environment. All red
supergiants were once blue supergiants (\S 1.1). Much speedier, $v_{\rm
wind}\sim10^3$ km s$^{-1}$, winds of blue supergiants exert vastly higher ram
pressure on the surrounding medium than the winds of red supergiants. On the
other hand, the fast wind would have quickly swept-up an equivalent amount of
interstellar matter and thus slowed-down significantly merely for reason of
conservation of momentum: this happens already within a century ---
$t=(3\dot{M})^{0.5}(4\pi\rho_{\rm ism})^{-0.5}v_{\rm wind}^{-1.5}$. Also, the
energetic radiation from the B-type star creates a Str\"omgren z\^one of warm
plasma, the internal pressure of which acts as a buffer against expansion of
the wind bubble. Before long, the wind bubble will not expand much faster than
a km s$^{-1}$. That said, the subsequent slow, dense red supergiant wind will,
at least initially, expand into a rarefied, and quickly-recombined,
circumstellar environment; this may alleviate the potential challenges for
such a wind to plough into a molecular cloud, for instance, and justify the
normally assumed density profile throughout the (outer) circumstellar envelope
of red supergiants, $\rho\propto r^{-2}$.

We must emphasize that mass lost in earlier phases {\it will} have an impact
on the properties of the red supergiant. This previous mass loss may be
modest, perhaps only amounting to a few per cent of the birth mass. However,
it is an uncertain quantity, and could be underestimated if significant
amounts of mass are lost in brief (and thus rarely witnessed) events such as
LBV-like eruptions.

\subsection{Implications of red supergiant mass loss}

Given the uncertainties in mass-loss rates and empirically-derived lifetimes,
the mass lost over the entire red supergiant phase, $\Delta
M=\int\dot{M}\,{\rm d}t$ is uncertain by an amount which is comparable to the
mantle mass. Red supergiants lose mass at a rate that is normally not higher
than the rate at which nuclear fuel is burnt, meaning that the star will enter
the next phase of evolution before having shed the entire mantle; however,
during periods of the most intense mass loss, at rates $\dot{M}\sim10^{-4}$
M$_\odot$ yr$^{-1}$, the mass loss greatly exceeds the nuclear consumption
(van Loon et al.\ 1999). The exact duration of that extreme phase is thus of
crucial importance; it definitely is brief, for few such examples are seen,
yet for that reason it is also difficult to estimate precisely. Furthermore,
the red supergiant lifetime may be extended by efficient mixing, for instance
as a result of rotation (Meynet et al.\ 2006); this means that the integrated
mass loss may differ between stars of the same birth mass (and even of the
same initial metallicity), further complicating theoretical predictions as
well as their empirical tests.

Although red supergiants with birth masses $<17$ M$_\odot$ are seen to explode
before having had time to shed their hydrogen-rich mantles, the fate of more
massive red supergiants is a mystery. At the top end, it is unclear whether
red supergiants in the approximate birth-mass range of 25--30 M$_\odot$
deplete their mantles enough to transform themselves into Wolf-Rayet stars.
Although it is difficult to establish direct links between Wolf-Rayet stars
and their red supergiant counterparts, the recently-determined, high,
empirical mass-loss rate estimates of red supergiants appear to be consistent
with the observed Wolf-Rayet star populations (Vanbeveren, Van Bever \& Belkus
2007). Post-red supergiant yellow hypergiants are extremely rare, though, but
examples have been suggested, {\it e.g..}, IRC+10\,420 (Kastner \& Weintraub
1995). At the bottom end of the mass scale, it is unclear whether super-AGB
stars shed their mantles to leave an oxygen-neon white dwarf or whether they
die prematurely; examples of each have been suggested, {\it e.g.}, by
Williams, Bolte \& K\"oster (2009) and Botticella et al.\ (2009),
respectively. As their luminosities and temperatures do not differ very much
from those of the most massive AGB stars, which do shed their mantles in time,
it is indeed possible that super-AGB stars may avoid explosion, too; on the
other hand, due to their different internal structure their atmospheres may
not become as loosely bound as those of AGB stars and thus their mass-loss
rates may not reach the same extreme values.

The slow, dense outflow from the red supergiant will be the stage for the next
act, be it the faster wind from a hotter star or the very fast blast wave and
ejecta from a supernova. A faster outflow will interact with the red
supergiant envelope, much as is believed to happen in planetary nebulae around
post-AGB objects. Equatorial density enhancements and faster bipolar outflows
are a common feature observed in red supergiants, for instance in their maser
profiles or imaged directly {\it via} their dust emission ({\it e.g.}, Ohnaka
et al.\ 2008). It is possible, therefore, that features such as the ring
observed in the SN\,1987A aftermath have their origin in the red supergiant
phase. Both blue-loop stars and post-red supergiant Wolf-Rayet stars (if they
exist) also develop a fast wind, running into a red supergiant envelope; hence
it may be possible to distinguish these post-red supergiant objects from stars
with similar spectral type evolving directly off of the main sequence, on the
basis of their circumstellar geometry --- at least in a statistical sense.

Finally, the relative amounts of mass in the envelope and mantle at the time
of explosion have a bearing on the amount of dust produced by red supergiants
(in the envelope, although this might be destroyed by the forward shock of the
supernova) and supernovae (in the reverse shock running through the ejected
mantle, although the dust-producing layers underly the hydrogen-rich mantle).
The mantle affects the explosive nucleosynthesis and mixing of its products,
as well as the luminosity and lightcurve of the supernova; the envelope
creates a light echo as it scatters the supernova's light, besides showing up
in the spectroscopic evolution of the supernova afterglow. These dependences
need yet to be quantified in reliable detail, but they show huge potential to
turn supernovae into probes of the red supergiant mass loss.

\acknowledgements I would like to express my thanks to the local and
scientific organisers for such an inspiring, pleasant, and smoothly-run
workshop. I am also grateful to the Royal Society for financial support of my
participation.

\end{document}